\documentclass[twocolumn,amsmath,amssymb]{revtex4}

\usepackage{graphicx}
\usepackage{bm}
\usepackage[T1,OT1]{fontenc}

\begin{document}

\title{Reshaping elastic nanotubes via self-assembly of surface-adhesive nanoparticles}

\author{Josep C. P\`amies}

\author{Angelo Cacciuto}\email{ac2822@columbia.edu}

\affiliation{Department of Chemistry, Columbia University, 3000 Broadway, New York, NY 10027}

\begin{abstract}
Elastic sheets with macroscopic dimensions are easy to deform by bending and stretching. Yet shaping nanometric sheets by mechanical manipulation is hard. Here we show that nanoparticle self-assembly could be used to this end. We demonstrate by Monte Carlo simulation that spherical nanoparticles adhering to the outer surface of an elastic nanotube can self-assemble into linear structures as a result of curvature-mediated interactions. We find that nanoparticles arrange into rings or helices on stretchable nanotubes, and as axial strings on nanotubes with high rigidity to stretching. These self-assembled structures are inextricably linked to a variety of deformed nanotube profiles, which can be controlled by tuning the concentration of nanoparticles, the nanoparticle-nanotube diameter ratio and the elastic properties of the nanotube. Our results open the possibility of designing nanoparticle-laden tubular nanostructures with tailored shapes, for potential applications in materials science and nanomedicine.
\end{abstract}


\maketitle

Direct mechanical manipulation can make macroscopic sheets conform to specific shapes. However, it is difficult to use mechanical means to reshape sheets with sizes in the micrometer range and smaller. An alternative at sub-micrometric scales would be the use of adhesive nanoparticles that induce local deformation and may catalyze global shape changes. Indeed, nanoparticles have been shown to drive the folding of graphene~\cite{graphene_folding}, of thin films of silicon~\cite{thin-film_folding} and carbon nanotubes~\cite{folding_carbon_nanotubes}, the budding of fluid membranes by protein aggregation~\cite{photosynthetic_membranes, aggregation_vesiculation}, and the deformation of vesicles via the adhesion of nanoparticles~\cite{nanoparticle_vesicle_adhesion, perling_vesicles_nanoparticles}.

When nanoparticles adhere and deform a surface, effective, curvature-mediated nanoparticle interactions arise as a result of the tendency of the surface to minimize the deformation caused by the nanoparticle imprints. One of the first studies of curvature-mediated interactions was that of Goulian and colleagues~\cite{curvature-mediated_interactions_1}, who calculated such effects in the context of protein aggregation in biological membranes. The effective pair interaction in fluid membranes is usually isotropic, has a Casimir-like functional dependence on particle separation~\cite{curvature-mediated_interactions_2} and a nontrivial dependence on the protein shape. However, for elastic (tethered) surfaces, the overall effect of the forces at play is more complicated. Unlike fluid surfaces, which cannot withstand shear, elastic thin sheets can stretch in response to the forces applied by strongly adhering nanoparticles. If the nanoparticles are able to diffuse on the sheet, they should be able to self-assemble in a configuration that reduces the mechanical cost of deforming the surface. However, the stretching energy associated with tethered surfaces imposes global geometric constraints to nanoparticle arrangements, and this leads to nontrivial many-body effects that extend across the surface. We expect the effective nanoparticle interactions to depend on the bending and stretching rigidities of the surface, its topology, and the relative location and specific extent of the local deformations imprinted by the nanoparticles. As a result, a significantly different --- and phenomenologically richer --- behavior emerges when nanoparticles aggregate on elastic surfaces rather than in fluid interfaces. Clearly, for multiple indenting (adhering) nanoparticles, one expects that the shape of the deformed surface is inextricably linked to the configuration of the adhered nanoparticles. Here, we use computer simulation to demonstrate that nanoparticles adhering and locally deforming the surface of an elastic nanotube can self-assemble into linear structures which themselves cause the global reshape of the nanotube. Our results suggest a new, nanoparticle-based route to design the effective elastic properties and the overall shape of a flexible surface.

To this aim, we considered a simple, coarse-grained model of an elastic nanotube with bending and stretching deformation modes, and of adhesive spherical nanoparticles (smaller than the tube radius) which adhere to the nanotube via a generic short-range potential acting between any nanoparticle and the surface. Monte Carlo simulations of this model show that, for a wide range of nanoparticle-nanotube diameter ratios, nanoparticles arrange into one-dimensional strings. Moreover, the strings acquire a preferential orientation which depends on the elastic rigidity of the nanotube: on stretchable nanotubes, strings of nanoparticles arrange as rings or helices; on nanotubes with high rigidity to stretching they form axial strings. We analyzed these arrangements as a function of bending rigidity, nanoparticle area-density (that is, the relative area of the surface covered by nanoparticles) and the average nanoparticle bound area. We also find that the cross-sectional profile of the nanotube can be shaped into ellipsoidal, triangular, rectangular and other regular and irregular forms, depending on the number and orientation of the self-assembled strings, the nanoparticle-nanotube diameter ratio and the elastic properties of the nanotube. Overall, our results suggest that nanoparticle organization and surface deformation can be controlled by tuning the concentration of adhesive nanoparticles and the mechanical properties of the surface.

\begin{figure}
    \includegraphics[height=3.0cm]{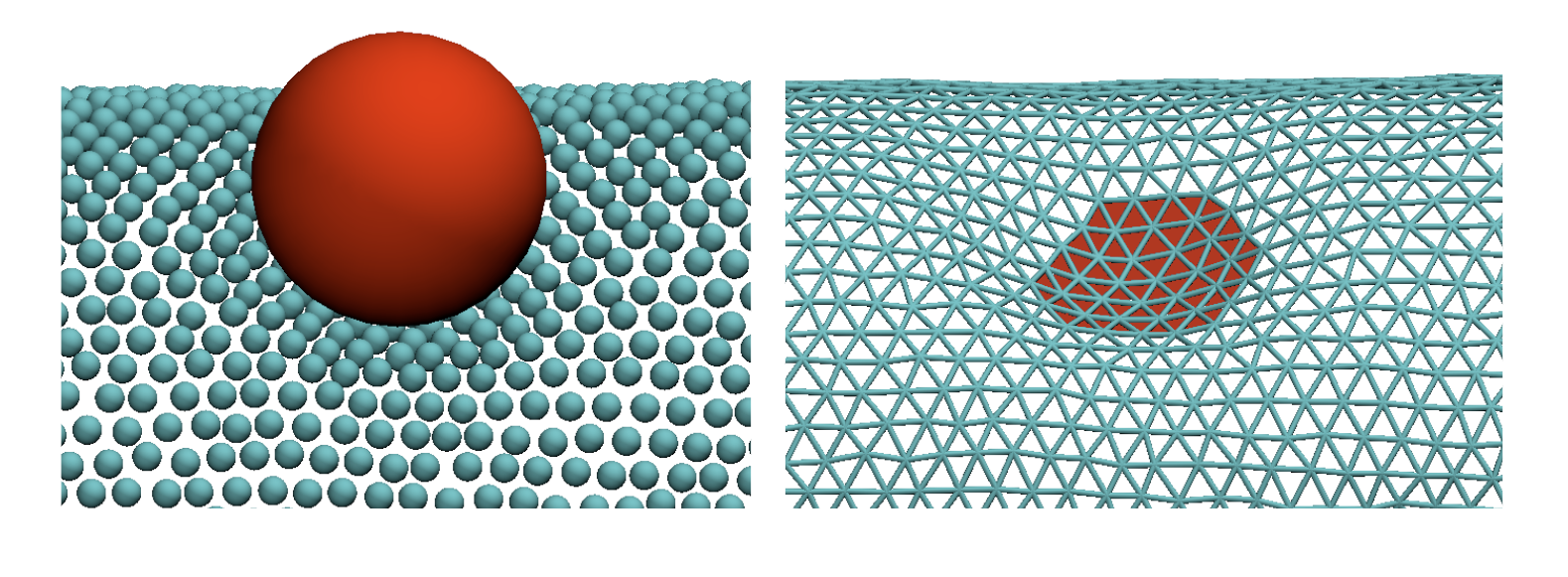}
    \caption{\label{fig:model}Detail of the particle (left) and triangulated-surface (right) representations of the nanoparticle-nanotube system. Nanotubes are made of hexagonally arranged beads of diameter $\sigma$. Nanoparticles have a diameter of $10\sigma$ and interact with the beads with energy per bead $\epsilon$ when a bead is within a distance $2\sigma$ from the nanoparticle surface. The nanoparticle bound area (bottom, shaded) is defined by the triangles whose all three vertices fall within the annular volume defined by the excluded volume of the particles and a sphere of radius $6.5\sigma$. We report the nanoparticle bound area as the average value, $A_b$, for all nanoparticles bound to the nanotube and as a percentage of the nanoparticle surface.}
\end{figure}

We modeled the elastic surface following a beads-and-string scheme~\cite{bead-and-spring_model}. The nanotube consists of a triangulated surface of hard beads connected by elastic links and arranged in a hexagonal lattice with an overall cylindrical shape, as sketched in Fig.~\ref{fig:model}. The triangulation of the network allows for an easy computation of elastic energies, with rigidities to bending and stretching, $\kappa$ and $\kappa_F$, respectively, characterizing the resistance to elastic deformation of the nanotube. The nanoparticles are described as hard spheres, and interact with the surface beads with a maximum energy per bead, $\epsilon$. Beads and nanoparticles do not interact with spheres of their own type, besides satisfying excluded volume constraints. We define the total energy of the model as the sum the Helfrich bending elasticity, the stretching energy of links, and a nanoparticle-surface adhesive potential, which are written as
\begin{eqnarray}
	\mathcal{H} & = & \sum_i \frac{3}{8} \kappa \frac{\left( \sum_j \lvert \, r_{ij} \, arccos(\textbf{n}_{ij} \cdot \textbf{n}_{ji}) \rvert \right)^2} {\sum_j A_{ij}} + \nonumber\\
	& + & \sum_{ij} -\frac{\kappa_F}{2} l_m^2 \ln\left(1 - [(r_{ij} - l_0) / l_m]^2 \right) + \\
	& + & \sum_{ik} \left\{
	\begin{array}{ll}
		\epsilon & \lambda_t \leq r_{ik} < \lambda_r \\
		\epsilon (\lambda^2_m - r_{ik}^2) / (\lambda^2_m - \lambda^2_r) & \lambda_r \le r_{ik} < \lambda_m \\
		0 & \lambda_m \le r_{ik}
	\end{array} \nonumber
	\right.,
	\label{eq:Hamiltonian}
\end{eqnarray}
The discretized bending term has been used before~\cite{photosynthetic_membranes, triangulation_method}, and is the sum of the bending energies of all triangle pairs of the surface. $r_{ij}$ is the length of the link connecting vertices $i$ and $j$; $\textbf{n}_{ij}$ and $\textbf{n}_{ji}$ are the unit normal vectors to the triangles sharing that link; and $\sum_j A_{ij}$ the sum of the areas of the triangles sharing vertex $i$. We describe the stretching contribution with a Finitely Extensible Nonlinear Elastic (FENE) potential~\cite{note_FENE}, where $l_0$ is the zero-energy length of a link and $l_m = 3\sigma$ its maximum length. We have chosen $l_0 = 1.229\sigma$, which corresponds to an area fraction for the beads of $60\%$~\cite{note_area_fraction}. This means that a nanoparticle of $10\sigma$ in diameter with $A_b = 5\%$ is bound to $15$ beads approximately. The adhesion term is a ramp well potential acting between each nanoparticle-bead pair, with a tangent distance $\lambda_t = 5.5\sigma$. For convenience we chose $\lambda_r = 6.5\sigma$ and $\lambda_m = 7.5\sigma$. We have seen that the specific shape of the adhesive term of the potential does not affect qualitatively our results, provided that the potential is short-ranged~\cite{note_shape_potential}.

Clearly, both elastic constants and the adhesion energy per bead affect the extent of the deformation. We find that a convenient geometric parameter to characterize deformation is the average nanoparticle area that is bound to the nanotube (see caption of Fig.~\ref{fig:model} for a precise definition). We simulate the nanotube with a periodic boundary in the direction parallel to its longitudinal axis $x$, and use periodic boundaries in the three cartesian coordinates for the nanoparticles. The initial configuration consists of an unperturbed nanotube, and of nanoparticles randomly placed in the simulation box. Monte Carlo moves at constant number of particles and temperature sample the configurational space by attempting changes in the position of vertices and nanoparticles. In order to allow for stretch-free configurations, changes in the length of the box parallel to the nanotube axis were also attempted, with constant pressure $P_x = 0$. The results presented in this work correspond to simulations with nanoparticles of $10\sigma$ in diameter. Additionally, we have simulated systems with larger, $15\sigma$-nanoparticles at various points in the parameter space, and have also compared our results to a similar simulation model~\cite{filament_deforming_nanotube}. System-size and surface-discretization effects do not appear to be significant. It is important to point out that the nanoparticle area-density affects self-assembly and deformation. Indeed, we observe that a nanoparticle area-density larger than $\approx0.2$ is necessary for the nanoparticles to aggregate into linear structures. As the area-density increases, the linear structures become progressively interconnected, approaching the limit of homogeneous nanoparticle coverage. In this study we have focused in the intermediate area-density regime, for which considerable deformation and a larger degree of nanotube shapes can be accessed.

\begin{figure}
    \includegraphics[scale=.3]{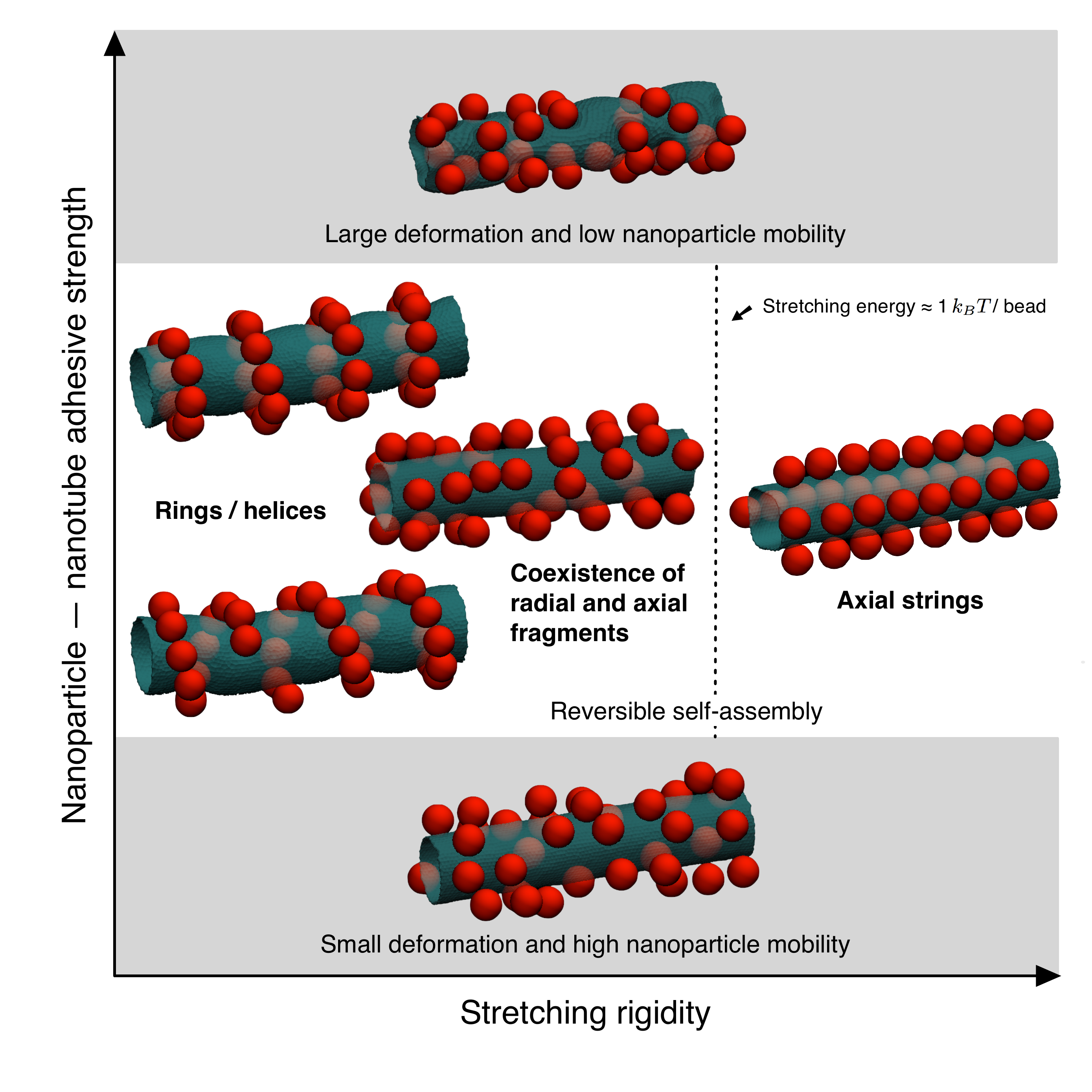}
    \caption{\label{fig:self-assembly} Nanoparticles self-assemble into ordered linear structures at moderate adhesive strength, switching orientation from rings and helices to axial strings when the stretching rigidity is such that the stretching energy is approximately $1\;k_{\rm B}T$ per bead. The boundaries of the simulated nanotubes match up to avoid boundary effects.}
\end{figure}

The average nanoparticle bound area controls the degree of deformation of the nanotube. For $20 \lesssim \kappa \lesssim 80\,k_{\rm B}T$ and $A_b \lesssim 5\%$, the elastic energy associated with the nanoparticle imprints is of the order of the thermal energy of the beads, and the nanoparticles can explore the entire surface on the nanotube. As $A_b$ increases, nanoparticles bound to the surface become less mobile. When $4\% \lesssim A_b \lesssim 10\%$, we observed that nanoparticles spontaneously organize into linear structures. For $A_b \gtrsim 10\%$, the adhesive energy needed to reach such bound area becomes larger than $10\,k_{\rm B}T$ per nanoparticle, and this makes the nanoparticle mobility on the surface too slow for the simulations to reach equilibrium within our simulation timeframe. This is schematically depicted in Fig.~\ref{fig:self-assembly}, where we also describe the self-assembled linear structures as a function of stretching rigidity. For fairly stretchable nanotubes, nanoparticles self-assemble into rings and helices, which can coexist on the same nanotube. For slightly less stretchable tubes, for which the stretching energy per bead is below $\approx 1\,k_{\rm B}T$, linear aggregates are shorter and their orientation can span all possible angles, although we have seen a preference for radial and axial orientations. Above that energy, we see preferentially axial arrangements.
\begin{figure}
    \includegraphics[scale=.3]{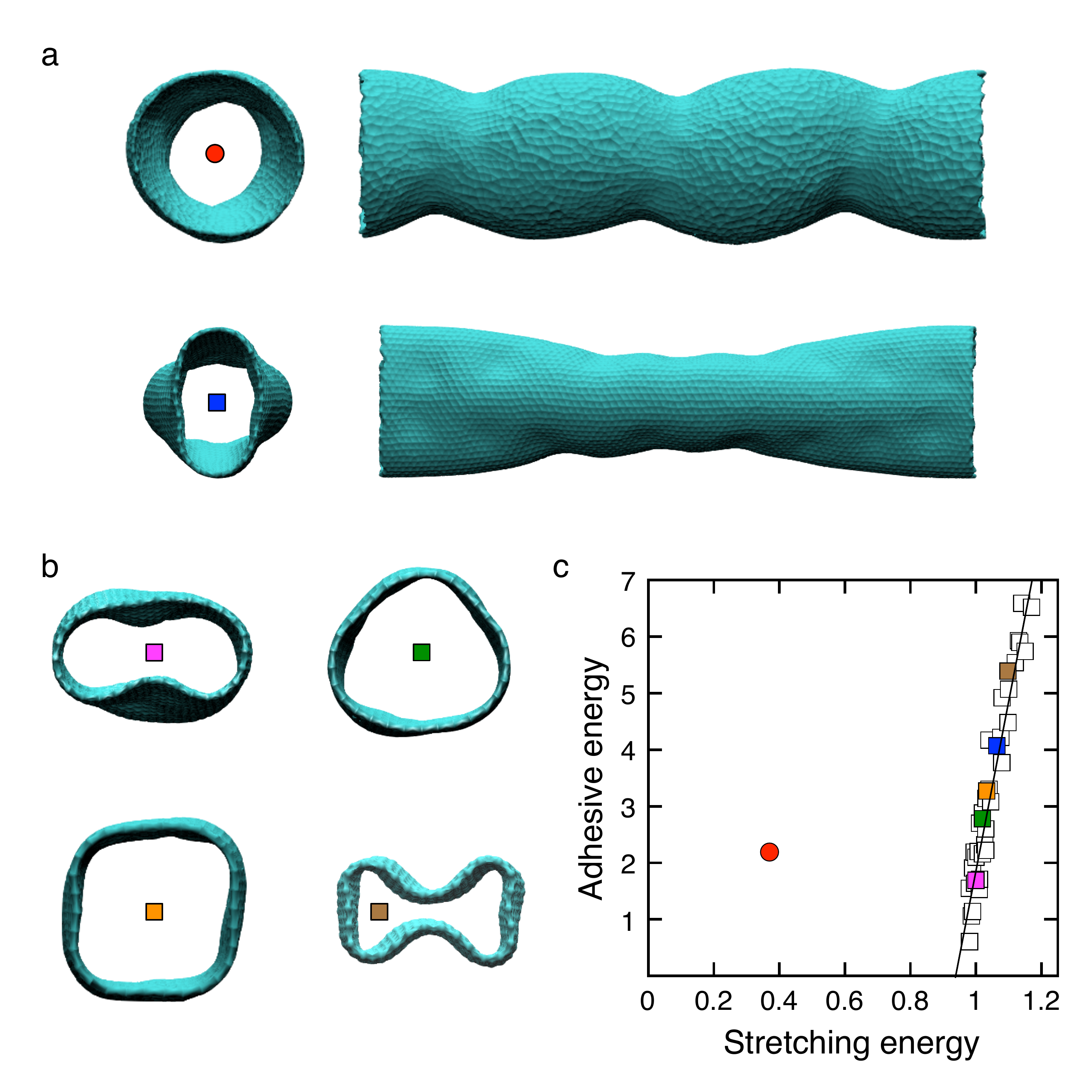}
    \caption{\label{fig:deformation} {\bf a,b}, cross-sectional cuts of deformed nanotubes. Symbols locate the corresponding energies in {\bf c}. Nanoparticles have not been drawn for clarity. {\bf c}, adhesion-stretching energies of configurations with axial strings (squares). One point corresponding to a helix (circle) is shown for comparison. Energies are given in units of $k_{\rm B}T$ per bead. The straight line is a fit to simulation data of axial strings with an (undeformed) nanotube-nanoparticle diameter ratio in the interval $1.017-2.033$, with $15-55$ nanoparticles, $A_b$ in the $5-8\%$ range, and bending and stretching rigidities within $20-100\;k_{\rm B}T$ and $120-960\;k_{\rm B}T/\sigma^2$.}
\end{figure}

Axial strings are essentially stretch-free configurations. Indeed, a nanotube completely rigid to stretching cannot accommodate nanoparticle rings or helices because these configurations necessarily involve double-curvature imprints, which have a stretching penalty. But the rigid nanotube can deform around axial strings by curving only in the radial direction. This is why axial strings only appear above $1\,k_{\rm B}T$ in stretching energy per bead. Indeed, there is a moderately sharp transition from the bending-dominated regime --- for which rings and helices are the bending-minimizing configurations --- to the stretching-dominated regime, in which the minimization of stretching energy becomes dominant. We believe that the fact that the transition between the two regimes takes place within a relatively small range of stretching energies has its origin in the high relative cost of stretching with respect to bending. Indeed, the stretching-to-bending energy ratio due to an indentation of depth $h$ in a thin shell of thickness $t$ can be written as $E_s/E_b \sim (h/t)^2 \sim \kappa_F / \kappa\,h^2$~\cite{Landau_Lifshitz}. This is also applicable to a nanotube with radius $R$ as long as $h \ll R$. Clearly, for sufficiently thin surfaces, the stretching energy cost overwhelms that of the bending, effectively imposing a global constraint on the possible deformations of the nanotube.

Next we discuss the shapes of the deformed nanotubes mediated by the linear aggregation of the nanoparticles bound to it, and also the connection between nanoparticle self-assembly and the shape of the nanotube. Fig.~\ref{fig:deformation}a shows two nanotube profiles: one with a helically shaped string of nanoparticles and another with four, almost-parallel axial nanoparticle strings. Ring and helical structures induce a screw-like type of deformation on the nanotube. Axial strings, however, induce a larger variety of shapes, which can also show constrictions along the nanotube axis if the strings are not fully parallel. Fig.~\ref{fig:deformation}b shows four different shapes corresponding to two (ellipsoidal profile), three (triangular profile) and four (square and bow-tie profiles) parallel axial strings. The more bent the nanotube profile is, the larger the adhesive energy per bead of the configuration, as can be seen in Fig.~\ref{fig:deformation}c. This figure also shows that stretching the surface above $1\,k_{\rm B}T$ per bead comes at a very steep cost in adhesion. This is largely due to the fact that the surface wrapping an axial string does not form an entirely straight channel, but follows the curvature of the individual nanoparticles forming the string.
\begin{figure}
    \includegraphics[scale=.3]{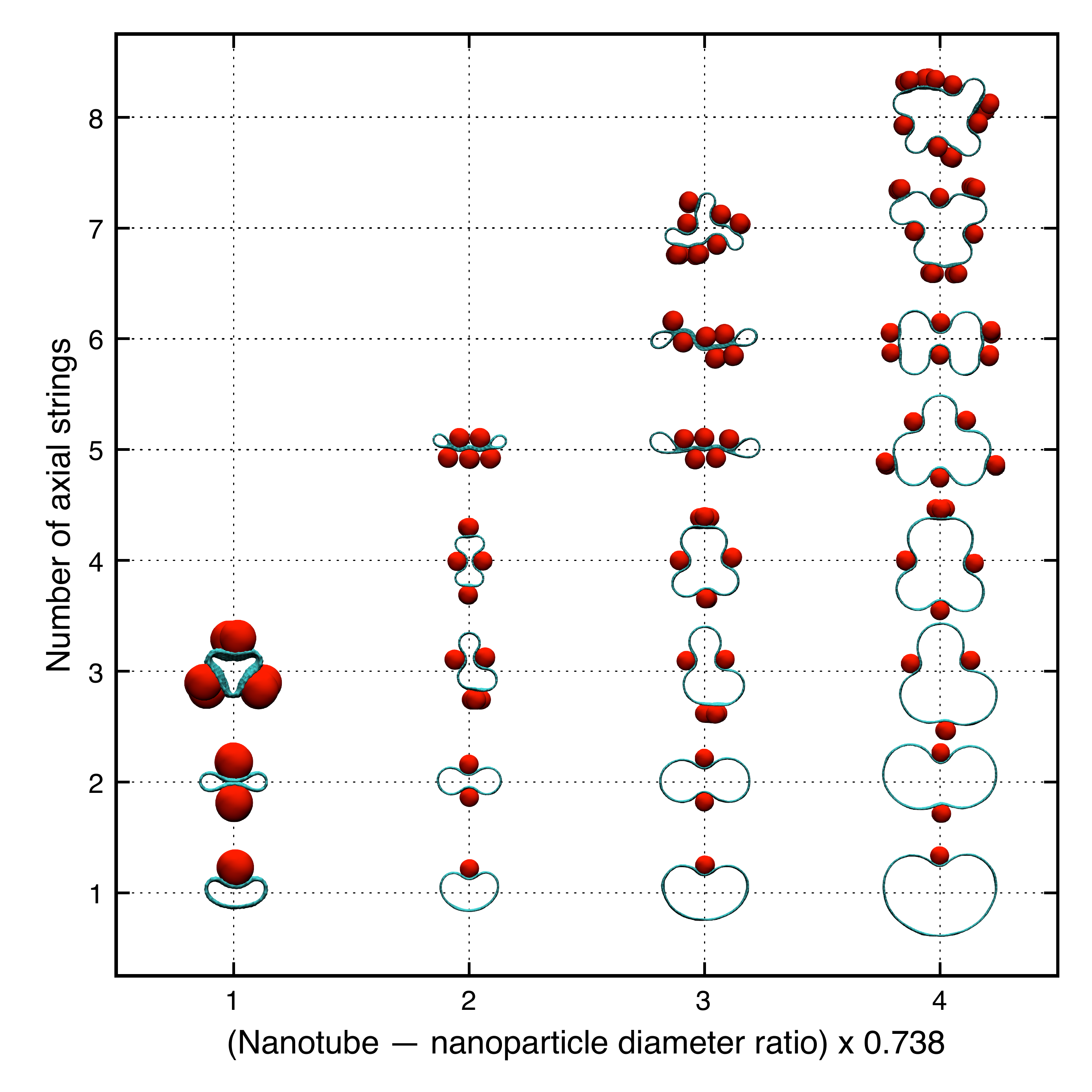}
    \caption{\label{fig:profiles} Snapshots of nanotube profiles with high rigidity to stretching, for which nanoparticles self-assemble into axial strings. The strings have been constrained to span the length of the tube and to contain the same number of nanoparticles each. The profiles thus have a uniform shape in the axial direction. Profiles in the leftmost column have been drawn twice as big for clarity. For all profiles the nanoparticle radius is $10\sigma$, $\kappa=40\;k_{\rm B}T$, $\kappa_F=500\;k_{\rm B}T / \sigma^2$ and $A_b = 10\%$.}
\end{figure}

Fig.~\ref{fig:deformation} suggests that, for the stretching-dominated regime, the shape of the nanotube can be controlled by the area-density of nanoparticles (or analogously, the number of axial strings) and the nanotube-nanoparticle diameter ratio. Fig.~\ref{fig:profiles} illustrates this by showing snapshots of the profiles obtained varying these two geometric parameters at constant bound area and elastic rigidities. We see that the profiles are fairly symmetrical but for the higher number of axial strings, for which the shape can presumably get trapped in metastable configurations. Also, a sufficiently high number of axial strings can cause a nanotube to collapse.

By altering the profile of the nanotube, one has direct access to the effective bending rigidity of the nanoparticle-nanotube composite in the axial direction, as such rigidity depends linearly on the tube's cross-sectional moment of inertia. Furthermore, it should be possible to reversibly switch nanoparticle self-assembly on and off in experiments --- for instance by tuning temperature, or by exploiting electrostatic or depletion interactions. Controlling nanotube shape may be relevant to applications of tubular nanomaterials~\cite{composite_nanotubes, polyelectrolyte_nanotubes, polymersomes_nanotubes}, of bioactive nanotubes~\cite{bioactive_polymer_nanotubes}, and in microfluidics~\cite{microfluidics}. We believe that experimental systems in which the here described coupling between nanoparticle self-assembly and nanotube deformation occurs can be readily realized. Also, our approach may inspire alternative routes to manipulating the folding of thin films of silicon for photovoltaic applications~\cite{thin-film_folding}.

We thank An{\fontencoding{T1}\selectfont\dj}ela \v{S}ari\'c and William L. Miller for stimulating discussions. This work was supported by the National Science Foundation under Career Grant No. DMR-0846426.


\begin{thebibliography}{21}
\bibitem[{not({\natexlab{a}})}]{note_FENE}
\bibinfo{note}{The FENE potential, most often used in bead-spring polymer
  models, is harmonic at its minimum and cannot be stretched beyond a maximum
  length $l_m$.}
\bibitem[{not({\natexlab{b}})}]{note_area_fraction}
\bibinfo{note}{A 60\% area fraction is a compromise between a sufficiently fine
  resolution of the triangulated surface and a reasonable fraction of accepted
  displacement moves.}
\bibitem[{not({\natexlab{c}})}]{note_shape_potential}
\bibinfo{note}{Compared to the simpler square-well potential, our particular
  choice improves the acceptance ratio of sampled states in the simulations.}
\end{thebibliography}
\end{document}